\documentclass[12pt]{iopart}
\usepackage{iopams} 
\usepackage[latin1]{inputenc}
\usepackage{graphicx}
\begin{document} 
 
\title[Effective non-retarded method as a tool for the design of
  optical properties]{Effective non-retarded method as a tool for the
  design of tunable nanoparticle composite absorbers}
\author{Guillermo Ortiz,$^{1}$ Marina Inchaussandague,$^2$ Diana
  Skigin,$^2$ Ricardo Depine,$^{2}$ and Luis Mochán$^{3}$}
\address{$^1$Departamento de F\'{\i}sica, Facultad de Ciencias
  Exactas, Naturales y Agrimensura, Universidad Nacional del
  Nordeste,\\ Avenida Libertad 5470, W3404AAS Corrientes, Argentina}
\address{$^2$Grupo de Electromagnetismo
  Aplicado, Departamento de F\'{\i}sica, FCEN, UBA \\ and IFIBA,
  CONICET, Pabellón 1, Ciudad Universitaria, 1428 Buenos Aires,
  Argentina} \address{$^3$Instituto de Ciencias F\'{\i}sicas,
  Universidad Nacional Aut\'onoma de M\'exico, \\ Apdo. Postal 48-3,
  62251 Cuernavaca, Morelos, M\'exico}
\ead{gortiz@unne.edu.ar}

\begin{abstract} 
We investigate the capabilities of the effective non-retarded method
(ENR) to explore and design nanoparticles composites with specific
optical properties. We consider a composite material comprising
periodically distributed metallic spheres in a dielectric host
matrix. The effective macroscopic dielectric function of the composite
medium is obtained by means of the ENR and is used to calculate the
electromagnetic response of a slab made of such an inhomogeneous
material. This response is compared with that obtained using the
Korringa-Kohn-Rostoker wave calculation method (KKR). We analyze the
optical properties for different filling fractions, especially in the
vicinity of the resonance frequencies of the macroscopic dielectric
function. We show that appropriately choosing the parameters of the
composite it is possible to achieve a tunable absorber film.  The ENR
results to be a versatile tool for the design of nanoparticle
composite materials with specific properties.
\end{abstract} 


\section{Introduction} 

The optical properties of inhomogeneous material systems have been
extensively studied for the last decades. The formulations developed,
such as the theories of spectral representations
\cite{Bergman(1976),Fuchs(1977),Milton(1980)}, spatial fluctuations
\cite{Mochan(1985)I,Mochan(1985)II}, renormalized polarizabilities
\cite{Barrera(1988),Fuchs(1991)} and diagrammatic series
\cite{Felderhof(1982),Barrera(1989)}, intended to explain the physical
phenomena involved in the electromagnetic macroscopic response of
these systems using analytical and/or semi-analytical expressions.
However, since the emergence and development of computational tools,
implementations such as the Discrete Dipole Approximation
\cite{Draine(1994),Khlebtsov(2000)}, and the Finite-difference
time-domain (FDTD) method \cite{Karkkainen(2000),Joannopoulos(2007)},
have gained widespread interest. The contribution of these
developments has been primarily focused on the corroboration of the
transmission/reflection spectra obtained experimentally in complex
systems. Modern technology possibilities of miniaturization and
self-assembling have accompanied these developments
\cite{Askari(2012),Hess(2012)}.

If the typical wavelengths are much larger than the size of the
inhomogeneities, it is possible to make a quasi-static treatment (also
called long wavelength approximation), in which many of the current
methods for determining the effective electric macroscopic
permittivity of the system are based. An enriching review of
traditional approaches to solve the problem by means of effective
medium models can be found in
\cite{Etopim(1977),Etopim(1993),Etopim(2002)}.  If the wavelength is
comparable to the size of the inhomogeneities, it is still possible to
use effective medium models to obtain the macroscopic dielectric
function of metal-dielectric composites
\cite{Silveirinha(2007),Huerta(2013)}, although in these cases the
dispersion of the dielectric response is both spatial and temporal.
We are interested in investigating inhomogeneous systems formed by two
ordered phases, one metallic and the other dielectric. The interest in
metal-dielectric composites lies in the possibility of designing their
optical properties by adapting the parameters of the materials. More
specifically, if the phases of the composite are chosen with
appropriate geometry and sizes, it is possible to tune the resonance
of the dielectric macroscopic function within the visible range
\cite{Ortiz(2009),Cortes(2010),Mochan(2010)}.

Effective medium theories describe the macroscopic dielectric function
of complex composite nanostructures in terms of the dielectric
functions of their components and of a limited number of geometrical
parameters. In the case of diluted systems with spherical inclusions,
the filling fraction is sufficient information to obtain the
macroscopic dielectric function, as established by the traditional
effective medium theories \cite{Bruggeman(1935),MaxwellGarnett(1904)}.
However, most of the interesting effects are more strongly manifested
for inclusions with complex geometries and close to the percolation
threshold of the conductive phase, i.e. in systems with high filling
fractions.  As a result, it raises the question of how these
formulations should be modified to account for the effects of more
complex and concentrated structures.  It is well known that when the
metallic particles which are close enough, multipolar interactions
among them give rise to a dielectric macroscopic response that
exhibits several resonances, regardless of the
inclusions'size~\cite{Doyle(1977),Claro(1984)}. However, following the
results by Waterman et. al.~\cite{Waterman(1986)}, many authors have
neglected the multipolar order contribution~\cite{Chern(2010)}, even
for filling fraction near the close packing
condition~\cite{Moroz(2009)}. This assumption has also been proposed
for the development of composite system with novel optical
properties~\cite{Yannopapas(2005)}.

Among the theories developed to study non-diluted
\cite{Barrera(1994),Ortiz(2003)b} and ordered
\cite{Claro(1984),Rojas(1986),Ortiz(2009)} systems, the effective
non-retarded method (ENR) has been recently proposed. This formalism
makes it possible to obtain the complex and frequency dependent
macroscopic dielectric function for arbitrarily shaped inclusions
periodically ordered in 2D or 3D arrays. It can also deal with
interpenetrating inclusions and dissipative and dispersive materials
\cite{Cortes(2010)}. Although the ENR allows dealing with inclusions
of arbitrary geometry, it is particularly suited for particles with
planar faces such as cubes or parallelepipeds. This is because its
numerical implementation requires a discrete representation of the
scatterers, for which a cubic grid is used to discretize the unit
cell. Given the demonstrated versatility of this method for predicting
the optical properties of complex materials \cite{Mochan(2010)}, it is
important to study its behavior when applied to the case of composites
with spherical inclusions, as those that have been recently
manufactured \cite{Geist(2005),Lee(2009),Muhlig(2011)}.  The case of
spherical inclusions requires a detailed analysis of the numerical
procedure employed to determine the dielectric macroscopic
function. Moreover, it is essential to validate the results by
comparison with rigorous methods which do not include approximations
in the representation of the shape of the inclusions.  Among these
methods, the KKR makes it possible to investigate the electromagnetic
response of composites formed by periodic arrays of spheres and was
shown to be numerically efficient. In the KKR, the electromagnetic
interactions between the scatterers are calculated by means of the
layer-multiple-scattering method for spherical particles
\cite{Modinos(1987),Stefanou(1998),Yannopapas(1999),Stefanou(2000)}. Recently,
simulated reflectance spectra calculated by means the vector KKR at
high order band frequencies, have shown a clear correlation between
theoretical and experimental results \cite{Dorado(2007)a}.  It is our
aim to study the applicability of the ENR to the case of composites
formed by spherical metallic inclusions embedded in a dielectric
medium. To do so, we investigate the electromagnetic response of a
composite slab and compare the results obtained using the ENR and the
KKR.  In Section \ref{resonancias}, the ENR is used to obtain the
macroscopic dielectric response of a medium comprising a simple cubic
lattice of spheres.  In Section \ref{slab}, the KKR is used to obtain
the electromagnetic response of a composite material slab and it is
compared with that given by the ENR. In Section \ref{Aplicaciones} we
explore the potential of the ENR as a design tool for tunable
absorbers from diluted to overlapping particles. Finally, concluding
remarks are provided in Section \ref{conclusiones}.

\section{The macroscopic dielectric function} \label{resonancias}

Let us consider a composite made up of arbitrarily shaped inclusions
arranged in a periodic 3D array. Provided that the length scale of the
inclusions in the material is small compared with the electromagnetic
wavelength, this mixture can be treated as a homogenized composite
material. The ENR permits calculating the frequency-dependent
macroscopic dielectric function of such composites.  It is based on
Haydock's recursive scheme \cite{Haydock(1980)} from which a
tridiagonal representation of a characteristic function
\cite{Cortes(2010),Mochan(2010)} allows to obtain the macroscopic
dielectric function ($\epsilon^M$) of the composite. It depends on the
shape of the inclusions, on the dielectric function of the components
and on the filling fraction ($f$) of the array (ratio of the volume
occupied by particles into the unit cell to the total volume of the
unit cell).  Since the coefficients that appear in the tridiagonal
representation depend on the shape of the particles and on $f$, but
not on the dielectric constant of the inclusions, once they are
obtained for a particular geometry and concentration of the scatterers
they can be used for inclusions of different materials, saving
computation time.
 
In what follows, we apply the ENR to investigate materials with
spherical inclusions arranged in a simple cubic lattice of period $L$.
Therefore, a unit cell of the composite material is a cube of side $L$
with the inclusion at its center
\begin{figure}
\centering
\includegraphics[width=0.8\textwidth]{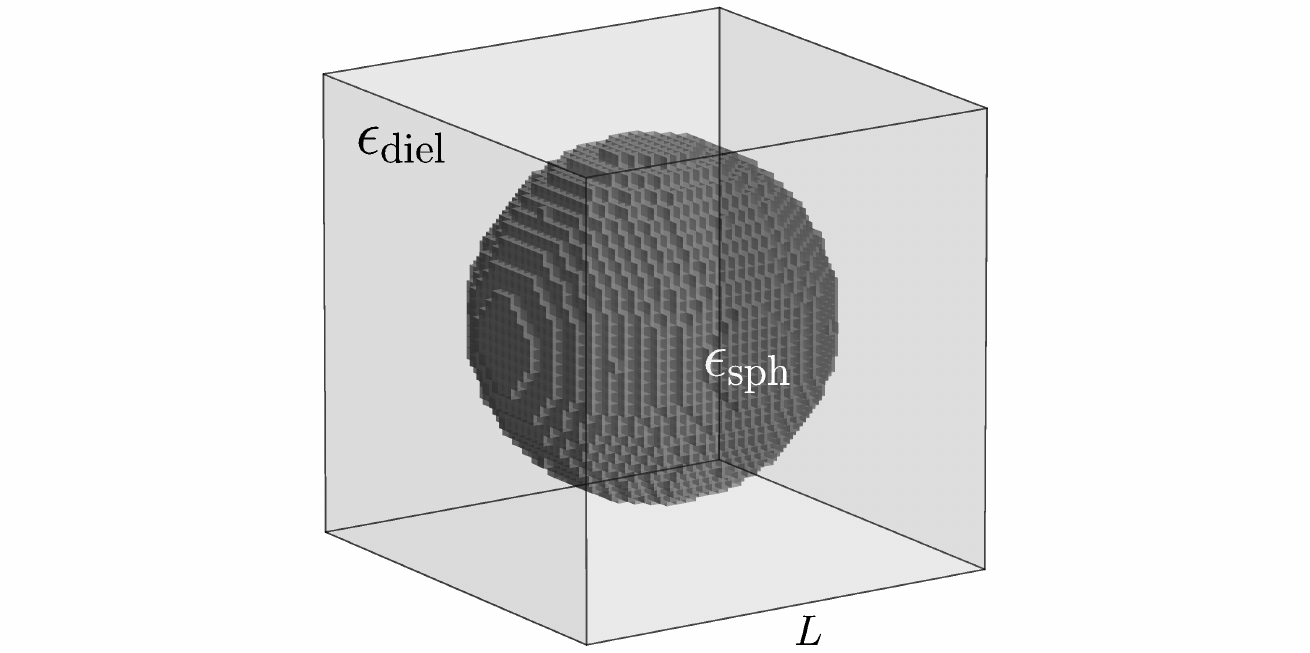}
\caption{ \label{fig2} Unit cell of a simple cubic lattice of
  spherical particles of radius $r=0.38L$ ($L$ being the side length
  of the unit cell) and dielectric function $\epsilon_{\rm sph}$
  embedded in a matrix with dielectric constant $\epsilon_{\rm diel}$.
  The sphere is discretized by small cubes of side $L/(2M+1)$ (in this
  scheme, $M=20$).}
\end{figure} 
(Fig. \ref{fig2}).  We consider an array of metallic spheres embedded
in a dielectric host material of permittivity $\epsilon_{\rm diel}$.
To model the permittivity of the metal $\epsilon_{\rm sph}$ as a
function of frequency $\omega$, we use the Drude formula
\begin{equation}
\epsilon_{\rm
  sph}(\omega)=1-\frac{\omega_p^2}{\omega^2+i\,\omega\,\Gamma}\,, \label{Drude}
\end{equation}
where $\omega_p$ is the bulk plasma frequency, $\Gamma$ stands for the
relaxation time and $i$ is the imaginary unit.
 
\begin{figure}
  \begin{center}
    \raisebox{6cm}{\Large
      a)}\includegraphics[width=.85\textwidth]{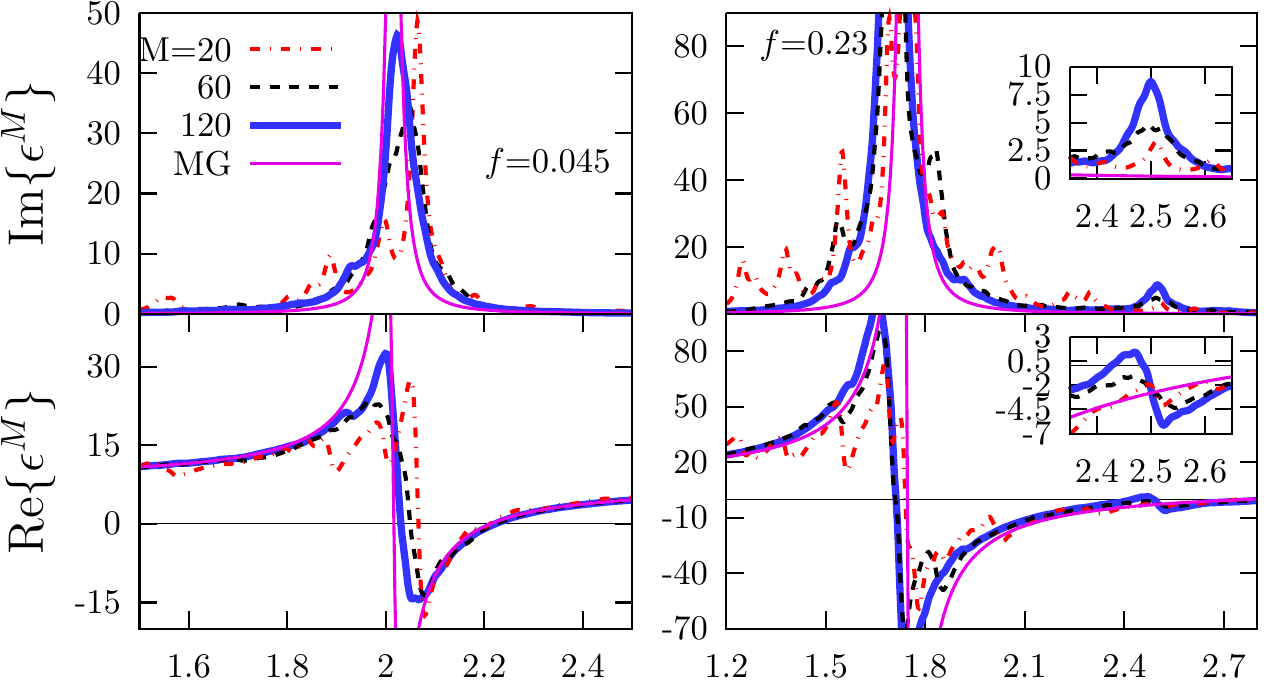}\\ \raisebox{6.5cm}{\Large
      b)}\includegraphics[width=.85\textwidth]{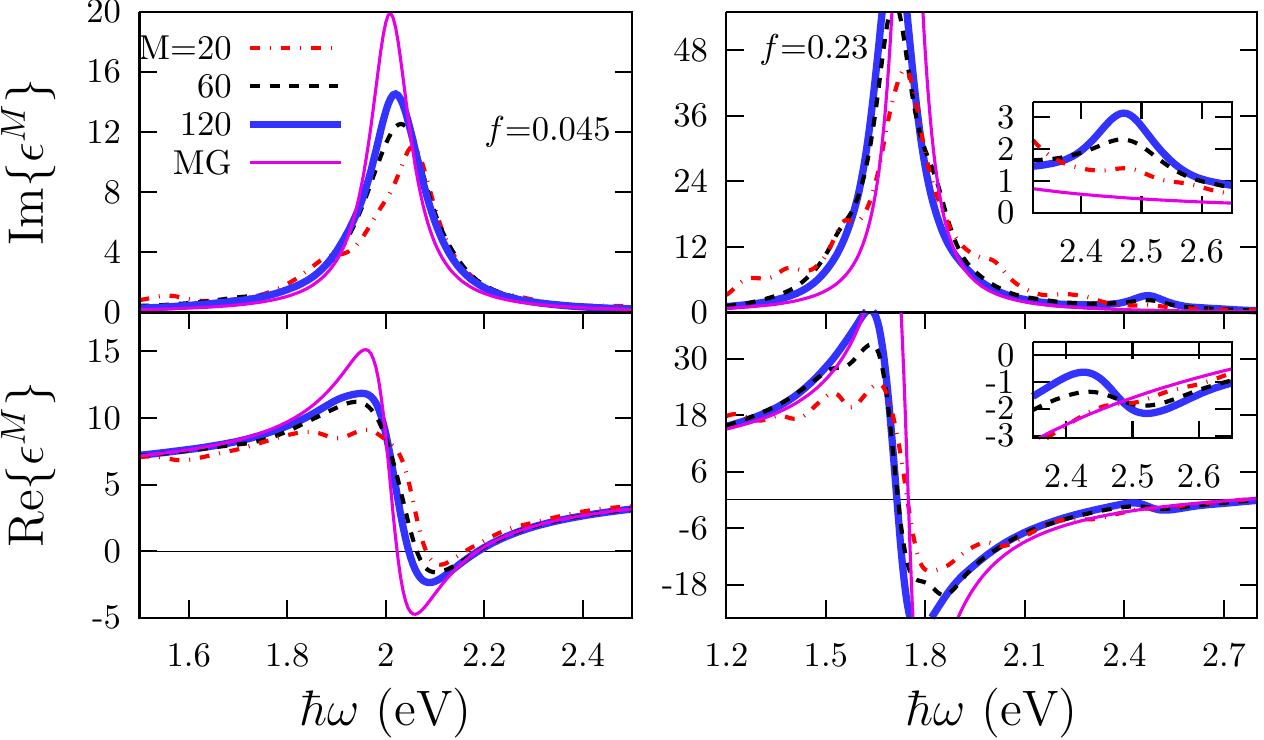}
    \caption{ \label{fig:rugo} Macroscopic dielectric function of a
      composite of metallic spheres embedded in a dielectric
      matrix. a) Silver spheres, $\epsilon_{\rm diel}=7.84$; b) gold
      spheres, $\epsilon_{\rm diel}=5.2$. Top (bottom) panel shows the
      imaginary (real) part. Metallic particles are arranged in a
      simple cubic lattice with $f=0.045$ for the left side and
      $f=0.23$ for the right one. Dashed-dotted lines, dashed lines,
      and thick solid lines correspond to $M=20$, 60 and 120,
      respectively.  For $f=0.23$, a detail of each curve in the
      vicinity of $\hbar\omega\approx$ 2.5 eV is displayed as an
      inset. Results given by the Maxwell-Garnett model are shown in
      thin solid lines.}
  \end{center}
\end{figure}

For the numerical implementation, the unit cell is divided into
$(2M+1)^3$ cubes of side $L/(2M+1)$, by taking $2M+1$ equidistant
points on each of its sides, and the discretized inclusion is defined
by the set of cubes contained within the volume of the actual
sphere. The discretization process is particularly suitable for
particles with planar facets, since they can be exactly represented by
stacking cubes.  For curved inclusions, the discretized particle
exhibits a stepped surface (as shown in Fig. \ref{fig2} for $M$ = 20)
that smooths by increasing $M$, i.e, by increasing the number of cubes
and thus by decreasing their individual volume.  The choice of $M$ is
not only important to properly represent the geometry of the particle
but also to obtain an acceptable resolution in the Fourier transform
calculations involved in the ENR~\cite{Cortes(2010)}.  The value of
$M$ that produces a valid description of the electromagnetic problem
must be found for each combination of parameters.  In
Fig.~\ref{fig:rugo} we show the real and imaginary parts of the
macroscopic dielectric function as a function of the photon energy
$\hbar\omega$ within the optical range for $M$ = 20, 60 and 120.  We
consider two different composites: one made of silver particles
($\hbar\Gamma=0.03$ eV and $\hbar\omega_p= 8.5$ eV) embedded in a
titanium dioxide matrix ($\epsilon_{\rm{diel}}=7.84$) (a), and the
other made of gold particles ($\hbar\Gamma=0.1$ eV and
$\hbar\omega_p=7.0$ eV) immersed in tellurium dioxide
($\epsilon_{\rm{diel}}=5.2$) (b). The values of $\hbar\omega_p$ and
$\hbar\Gamma$ considered in the examples were obtained by fitting the
real and imaginary parts of the dielectric function given by
(\ref{Drude}) to the experimental data of Ref.\cite{Johnson(1972)} in
the NIR-VIS range. Two different filling fractions are considered in
both cases: $f$ = 0.045 (left panels) and $f$ = 0.23 (right
panels). For a simple cubic lattice, $f=(4/3)\pi(r/L)^3$, and
therefore, $r/L=0.22$ for $f=0.045$ and $r/L=0.38$ for $f=0.23$. In
the figure, we also include the curves obtained by means of the
Maxwell-Garnett approach (MG).  As shown in Fig. \ref{fig:rugo}(a),
the curves exhibit a sort of ripple that smooths as $M$ is increased.
This fluctuation is more significant for the composite of silver
spheres than for that of gold spheres (Fig. \ref{fig:rugo}(b)),
suggesting that the choice of a proper value of $M$ is more critical
when the contrast between the permittivities of the constituent
materials is higher and also for smaller values of $\Gamma$.  Since
the memory requirements for computing the coefficients of the
tridiagonal representation of the characteristic function of the
composite increase with $(2M +1)^3$, in what follows we have chosen $M
= 120$, which is the maximum allowable value of $M$ for serial
calculations~\cite{pdl2010}.  This choice seems to be appropriate from
the point of view of the first element of the tridiagonal matrix:
since this coefficient is the average of the characteristic function
within the unit cell~\cite{Mochan(2010)}, its value should tend to $f$
as $M$ increases. Taking $M=120$ the relative error of this
coefficient is $\approx$ 0.04\% for $f = 0.045$.  Since for a fixed
value of $M$ unit cells with larger $f$ are better represented than
those with smaller $f$, a relative error smaller than 0.04\% is
guaranteed for $f \ge 0.045$.
 
From Mie theory, an isolated sphere is expected to exhibit multipolar
resonances for $\epsilon_{\rm sph} = -\epsilon_{\rm diel}(l+1)/l$,
where $l$ is a positive integer \cite{Kreibig(1995)}.  For
$\epsilon_{\rm sph}$ given by Eq. (\ref{Drude}), the resonant
frequencies are
\begin{equation} 
\omega \approx \sqrt{\omega_p^2/(1+\epsilon_{\rm diel}
  (l+1)/l)}.  \label{eq:resonancias}
\end{equation} 
For the composite system, these resonances can be visualized as peaks
in the imaginary part of the macroscopic dielectric function.  For
$f=0.045$, the curve of $\mbox{Im}\, \{\epsilon^M\}$ exhibits a peak
close to $\hbar \omega$ = 2 eV, which for $f=0.23$ appears at $\approx
1.71$ eV. These peaks are associated to the dipole resonance ($l=1$),
for which equation (\ref{eq:resonancias}) predicts a resonant value at
$\hbar \omega \approx 2.08$ eV for both composites considered in
Fig. \ref{fig:rugo}. As expected, this value is in very good agreement
with the spectral position of the peak for the diluted system
($f=0.045$).  However, for $f=0.23$ the multiple scattering between
the spheres has a significant impact on the optical response of the
system and consequently, the resonant frequency is shifted with
respect to the location predicted by Eq. (\ref{eq:resonancias}) for an
isolated sphere.  In this case, $\mbox{Im}\, \{\epsilon^M\}$ exhibits
also another peak at $\approx$ 2.5 eV which is associated to an
octupole resonance ($l=3$ in Eq. (\ref{eq:resonancias}))
\cite{Claro(1984)}. Notice that only resonances with odd $l$ can be
excited due to the system symetry.  This peak does not appear in the
MG curves since this approach neglects higher order interactions
between the scatterers, and therefore, it is only suitable for
treating diluted systems. The overall matching of the ENR and MG
curves is very good.  However, the dipole resonance peak obtained by
the ENR is lower and slightly wider than that predicted by the MG and
this can be explained by taking into account the discretized nature of
the spherical particle within the ENR.  In all cases, $\mbox{Re}
\,\{\epsilon^M\}$ and $\mbox{Im}\,\{\epsilon^M\}$ satisfy the
Kramers-Kroning relationship, as expected.\\

\section{Optical properties of nanocomposite thin films}\label{slab}
 
\begin{figure}
  \begin{center}
    \includegraphics[trim= 1cm 8cm 0 1cm,width=1.1\textwidth]{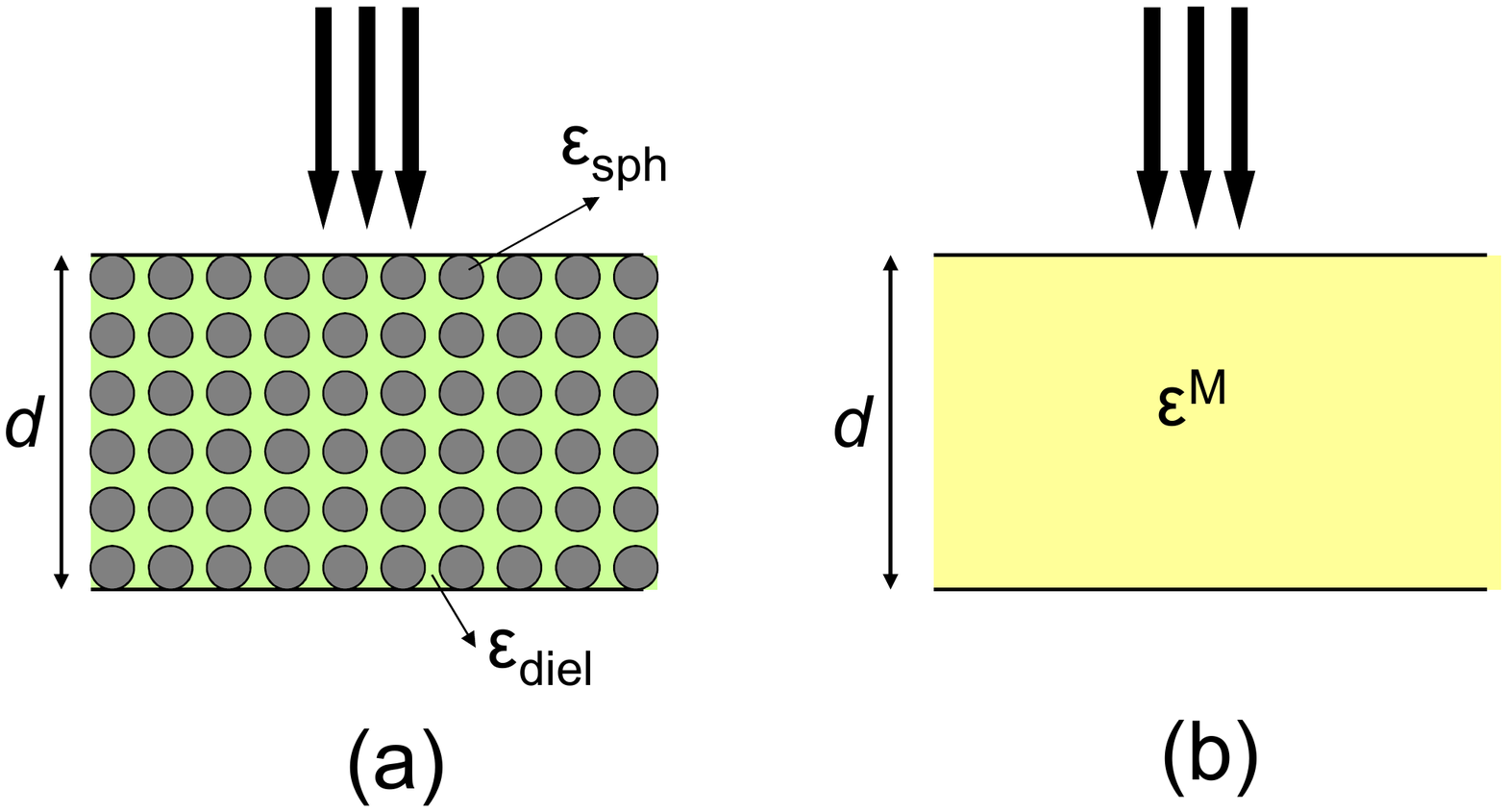}
  \end{center}
\caption{\label{fig:esquema} Scheme of the nanocomposite free standing
  film (thickness $d$).  The black arrows indicate the direction of
  the incident field.  (a) KKR: composite slab of metallic spheres of
  permittivity $\epsilon_{\rm sph}$ immersed in a dielectric host
  medium of permittivity $\epsilon_{\rm diel}$. In the figure, the
  number of layers of spheres is $N$ = 6.  (b) ENR: slab of a
  homogeneous material characterized by the macroscopic dielectric
  function $\epsilon^{M}$. }
\end{figure}

In this section we analyze the electromagnetic response of a slab made
of a composite material.  We first use the KKR to investigate the
influence of the number of layers of spheres in its optical properties
and establish the conditions for which the results given by the KKR
permit a valid comparison with those of the effective medium theory
(ENR). In Fig. \ref{fig:esquema} we show the scheme of the composite
slab modeled with the KKR (a) and with the ENR (b).
 
\subsection{The KKR method}\label{KKR}

As stated in the Introduction, among the methods available for the
calculation of the electromagnetic response of composite periodic
structures made of spheres, the KKR appears to be numerically
efficient. Within its framework, the electromagnetic interactions
between the inclusions arranged in the periodic lattice are calculated
by means of the layer-multiple scattering method for spherical
scatterers
\cite{Modinos(1987),Stefanou(1998),Yannopapas(1999),Stefanou(2000)}. The
crystal can be considered as a stack of parallel layers formed by
spheres periodically arranged in a 2D Bravais lattice. To solve the
electromagnetic problem, the multiple scattering between spheres of
each single layer is calculated first. Then, the scattered response of
multiple layers is determined by using a procedure similar to the one
used to calculate the reflection and transmission properties of
stratified media with planar interfaces.  The computer program MULTEM
\cite{Stefanou(1998),Stefanou(2000)} is a numerical implementation of
the KKR.
 
As mentioned in Section \ref{resonancias}, to compute the optical
response by the ENR is enough to set the filling fraction of the
structure. On the other hand, to perform the simulations using MULTEM,
the number of layers of spheres within the slab ($N$) is also
required.  As a consequence, if we fix the slab thickness ($d$) and
the filling fraction of the nanomaterial, the lattice parameter and
the radius of the spheres are determined by the relations $L=d/N$ and
$r=L(3f/4\pi)^{1/3}$ (for a simple cubic lattice of non-overlapping
spheres), respectively. This is a very important aspect to remark, as
it implies that for a given $d$, by varying $N$, it is possible to find
different pairs of values $r$, $L$ that give the same $f$.

\begin{figure}
  \begin{center}
    \raisebox{7cm}{\Large a)}\includegraphics[width=0.45\textwidth]{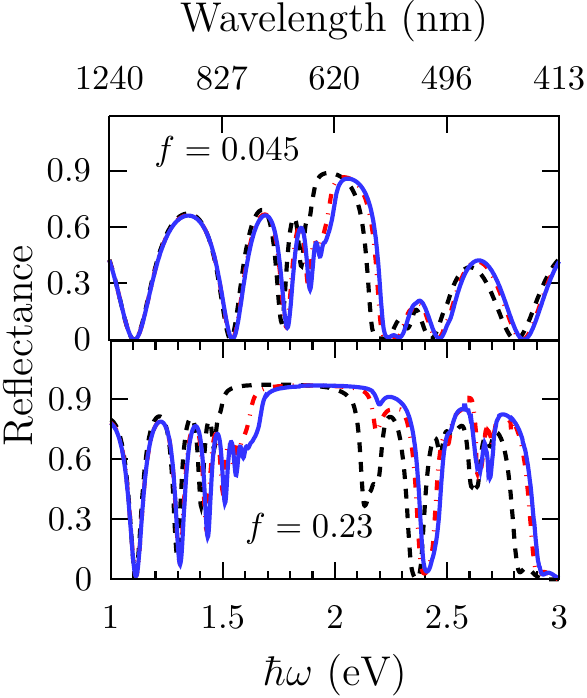}\includegraphics[width=0.45\textwidth]{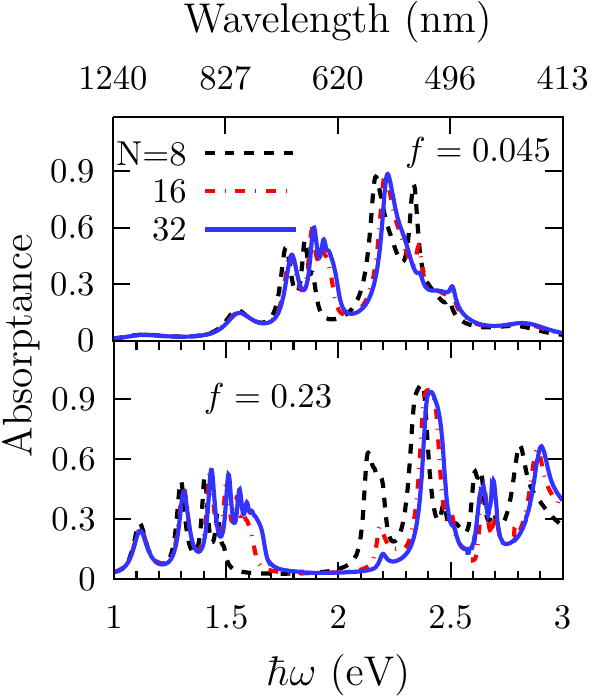}\\[1cm]
\raisebox{7cm}{\Large b)}\includegraphics[width=0.45\textwidth]{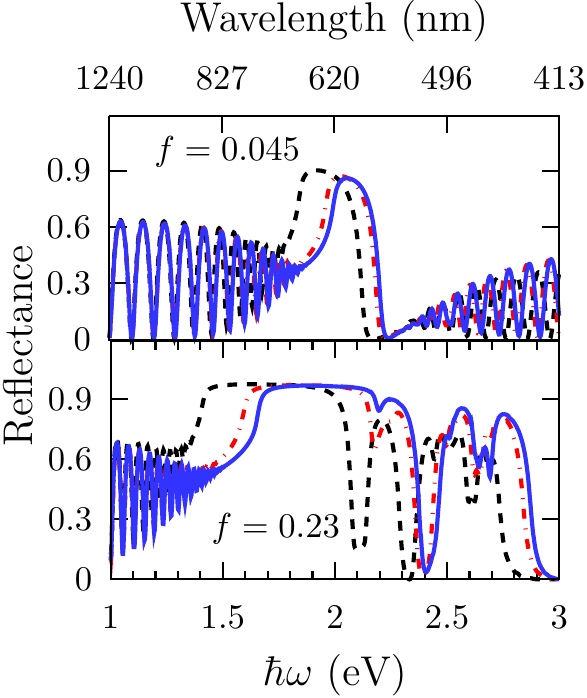}\includegraphics[width=0.45\textwidth]{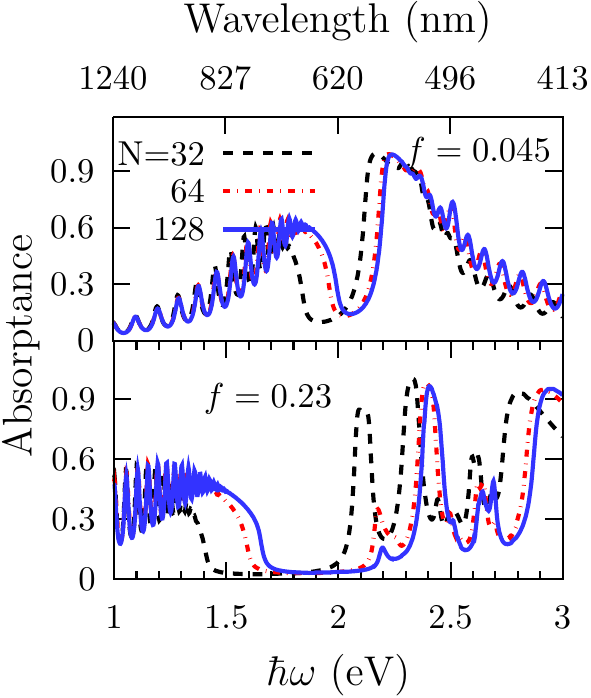}
    \caption{ \label{fig:RyAvsN} Reflectance (left panel) and
      absorptance (right panel) of a composite slab made of silver
      spheres embedded in a dielectric matrix of $\epsilon_{\rm
        diel}=7.84$. a) $d=360$ nm (dashed, dot-dashed and solid
      lines, correspond to $N=$8, 16 and 32, respectively); b) $d=1820$
      nm (dashed, dot-dashed and solid lines, correspond to $N=$32, 64
      and 128, respectively).  Top (bottom) panels correspond to
      $f=0.045$ ($f=0.23$).}
  \end{center}
\end{figure}

To investigate the influence of the number of layers for fixed values
of $d$ and $f$, we plot the reflectance and the absorptance
(Fig. \ref{fig:RyAvsN}) as a function of the photon energy
$\hbar\omega$ for $d=360$ nm and 1820 nm. The different curves in each
panel correspond to different values of $N$. For both thicknesses, we
consider two different filling fractions, $f=0.045$ (top panel) and
$f=0.23$ (bottom panel).

With increasing $N$ (for fixed values of $f$ and $d$), the radius of
the spheres becomes smaller, the wavelength-to-radius ratio increases,
and the long wavelength condition (in which the material behaves as a
homogeneous medium) is achieved.  Then, it is to expect that for
increasing $N$ the curves tend to a limit response given by the long
wavelength condition. This behavior can be observed in
Fig. \ref{fig:RyAvsN}.  In Table \ref{table:multem} we summarize the
values of $L$ and $r$ for the slabs investigated in the present
example.  According to the values of $r$ given in Table
\ref{table:multem}, we can conclude that 32 layers for $d$ = 360 nm
and 128 layers for $d$ = 1820 nm are sufficient to ensure the long
wavelength condition within the spectral range considered.
 
\begin{table}[h] 
\caption{\label{table:multem} Geometrical parameters of the composite
  slab considered in Fig. \ref{fig:RyAvsN}.}
\begin{center}
\begin{tabular}{|c|c|c|c|c|}\hline
$d$(nm) &$N$ & $L$(nm) & $r$(nm) ($f=0.045$)& $r$(nm) ($f=0.23$)\\ \hline 
& 8  & 45    &  9.9  & 17.1  \\ 
360 & 16 & 22.5  &  4.95 & 8.55  \\  
& 32 & 11.25 &  2.47 & 4.27 \\ 
\hline
& 32 & 56.87   &  12.51  & 21.61 \\ 
1820 & 64 & 28.44   &  6.25   & 10.80 \\
& 128 & 14.22   &  3.14   & 5.41 \\
\hline
\end{tabular}
\end{center} 
\end{table}

\subsection{Comparison between the KKR and the ENR}\label{comparacion}  
 
\begin{figure}
  \begin{center}
    \raisebox{6cm}{\Large a)}\includegraphics[width=0.3\textwidth]{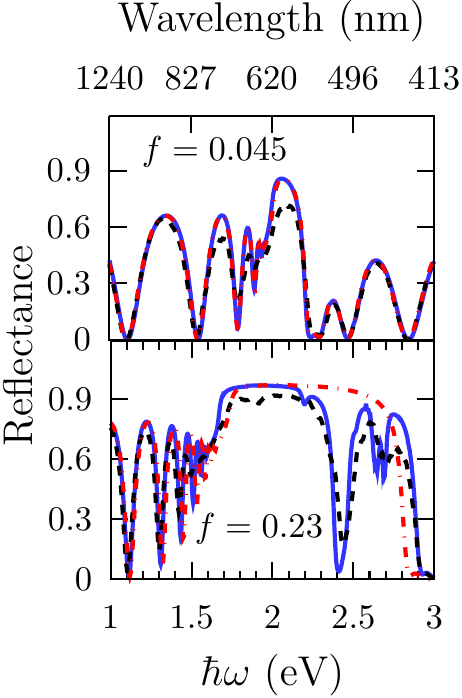}\includegraphics[width=0.3\textwidth]{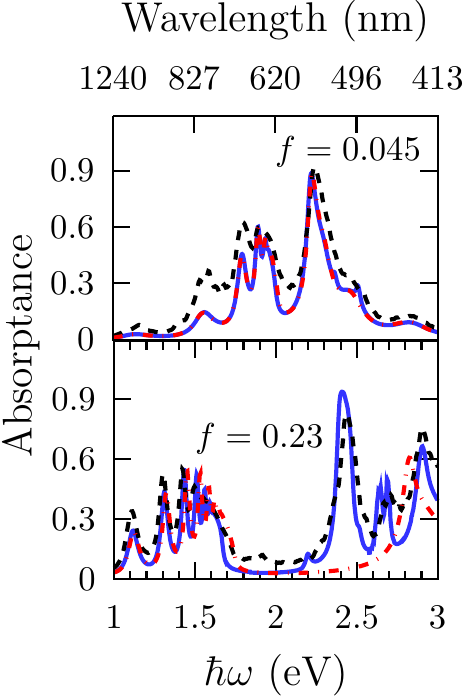}\includegraphics[width=0.3\textwidth]{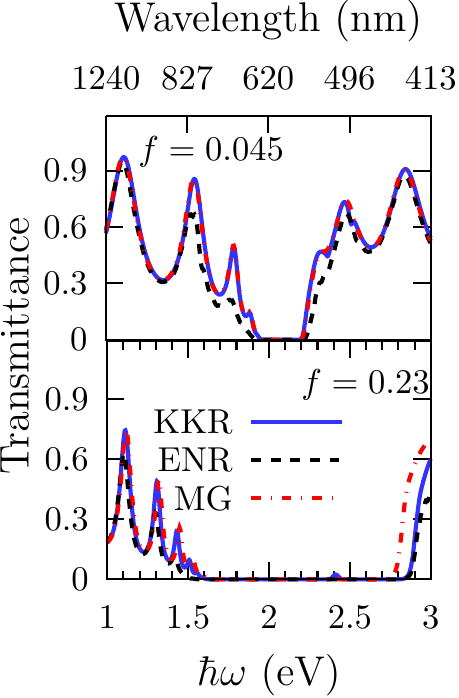}\\[1cm]
    \raisebox{6cm}{\Large b)}\includegraphics[width=0.3\textwidth]{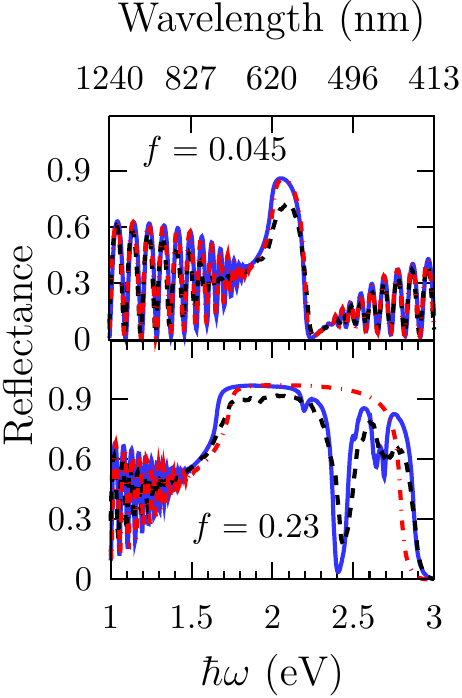}\includegraphics[width=0.3\textwidth]{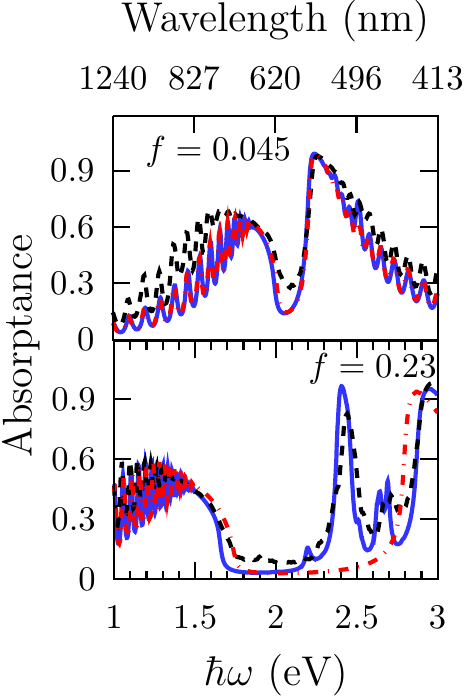}\includegraphics[width=0.3\textwidth]{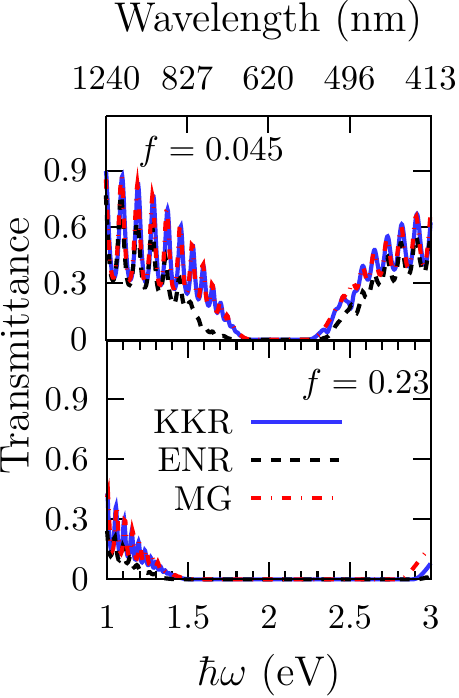}
    \caption{ \label{fig:RAyTvsMethods} Reflectance (left panel),
      absorptance (center panel) and transmittance (right panels ) of
      a composite slab made of silver spheres embedded in a dielectric
      matrix of $\epsilon_{\rm diel}=7.84$. a) $d=360$ nm and b)
      $d=1820$ nm.  Top (bottom) panels correspond to $f=0.045$
      ($f=0.23$). Solid, dashed, and dot-dashed lines correspond to
      KKR, ENR, and MG, respectively.}
  \end{center}
\end{figure} 

We now consider a plane wave normally incident on a free standing film
made of a composite medium of silver particles embedded in titanium
dioxide, as that considered in Fig. \ref{fig:rugo}(a).  Figure
\ref{fig:RAyTvsMethods} display the reflectance, absorptance and
transmittance for two filling fractions, $f$ = 0.045 (top panel) and
$f$ = 0.23 (bottom panel), and for $d$ = 360 nm (a) and 1820 nm
(b). We present curves obtained using three different approaches: KKR,
ENR and MG.  For all the ENR calculations we used $M$ = 120 . Taking
into account the analysis of Section \ref{KKR}, for the KKR
simulations we used $N$ = 32 for $d$ = 360 nm and $N$ = 128 for $d$ =
1820 nm.  The electromagnetic response of the slab is closely linked
to the macroscopic dielectric function of the composite, i.e., to the
behavior of $\mbox{Re} \,\{\epsilon^M\}$ and $\mbox{Im}
\,\{\epsilon^M\}$, which are strongly frecuency dependent (see
Fig. \ref{fig:rugo}). According to this, the response of the slab
exhibits very different features in the different spectral regions. At
low frequencies, the curves for both values of $f$ present
oscillations due to the Fabry-Perot interference and, as expected, for
the thicker slab the adjacent maxima of these oscillations are closer
to each other. In this spectral region ($\hbar \omega <$ 2 eV for $f$
= 0.045 and $\hbar \omega <$ 1.7 eV for $f$ = 0.23), $\mbox{Re}
\,\{\epsilon^M\} >0$ (as shown in Fig.  \ref{fig:rugo} (a)) and
therefore, the effective medium behaves as a lossy transparent
material that allows transmission through the slab.  However, when
approaching the dipole resonance frequency from the low frequency
side, $\mbox{Re} \,\{\epsilon^M\}$ and $\mbox{Im} \,\{\epsilon^M\}$
increase, and as a result, a lower transmittance is observed.
This behavior becomes more evident for the thicker slab as observed in
the right panels of Fig. \ref{fig:RAyTvsMethods} (b), in the ranges 1.8
eV $<\hbar \omega <$ 2 eV for $f$ = 0.045 and 1.5 eV $<\hbar \omega <$
1.7 eV for $f$ = 0.23.  It can be observed that for both thicknesses
considered, there is a reflectance band that starts at the dipole
resonance frequency: at $\approx 2$ eV (620 nm) for $f$ = 0.045 and at
$\approx 1.7$ eV (729.4 nm) for $f$ = 0.23 (see left panels of
Fig. \ref{fig:RAyTvsMethods}).  Within this band, the reflectance is
very high (more than 80\%) and the transmittance is negligible.
Notice that at the dipole resonance $\mbox{Re} \,\{\epsilon^M\}$
changes its sign from positive to negative and stays negative in a
frequency range that defines the reflectance bandwidth. This band is
wider for $f = $ 0.23 (1.7 eV $<\hbar \omega <$ 2.9 eV) than for $ f =
$ 0.045 (2 eV $<\hbar \omega <$ 2.2 eV) because the dipolar
interaction is stronger for higher concentrations of spheres, as
observed in Fig. \ref{fig:rugo}(a). Besides, for $f$ = 0.23 there is a
pronounced dip within the reflectance band very close to a multipole
resonance of higher order, which occurs at $\approx$ 2.5 eV. This dip
is located within a narrow frequency range at which $\mbox{Re}
\,\{\epsilon^M\}$ is positive (see Fig. \ref{fig:rugo} (a)) and in
this range the effective medium behaves as a lossy transparent
material which allows light propagation. However, the electromagnetic
response in this region differs substantially from that exhibited in
the low frequency zone, in which the condition $\mbox{Re}
\,\{\epsilon^M\}>0$ also holds. In the vicinity of the multipole
resonance, $ \mbox{Re} \,\{\epsilon^M\} \approx 1$ and therefore, the
contrast between the macroscopic permittivity of the slab and that of
vacuum is very low.  Consequently, the reflectance becomes negligible
and most of the incident energy enters the slab and is largely
absorbed, although in this region $ \mbox{Im} \,\{\epsilon^M\}$ is not
as high as in the dipole resonance. This effect has already been
reported in 2D~\cite{Ortiz(2009)} and 3D structures~\cite{Mochan(2010)}.
  
In general, the qualitative agreement between the ENR and the KKR
curves is satisfactory.  Both methods adequately describe the spectral
response of the composite slab. However, there are some quantitative
differences in certain spectral ranges which could be due to the
discretization of the sphere in the ENR. The surface roughness of the
discretized particles produces modes that affect the electromagnetic
response.  For $f$ = 0.045, the KKR and the MG curves are in excellent
agreement, as expected for a diluted system of spheres. However, for
$f$ = 0.23, the MG does not predict the splitting of the reflectance
band because it does not take into account multipolar interactions of
higher orders. Conversely, since the KKR and the ENR consider in their
formulations multipolar interactions between the spheres, they
adequately describe the splitting of the reflection band at those
frequencies at which multipole resonances are excited.

From the point of view of potential applications, compound metal-dielectric  
films can be used as tunable
frequency light absorbers~\cite{Ghenuche(2012)} by appropriately
choosing $f$ and $\epsilon_{\rm diel}$.

\section{Application example: tuning of the absorption bands} \label{Aplicaciones}
  
\begin{figure}
  \begin{center}
    \includegraphics[width=0.9\textwidth]{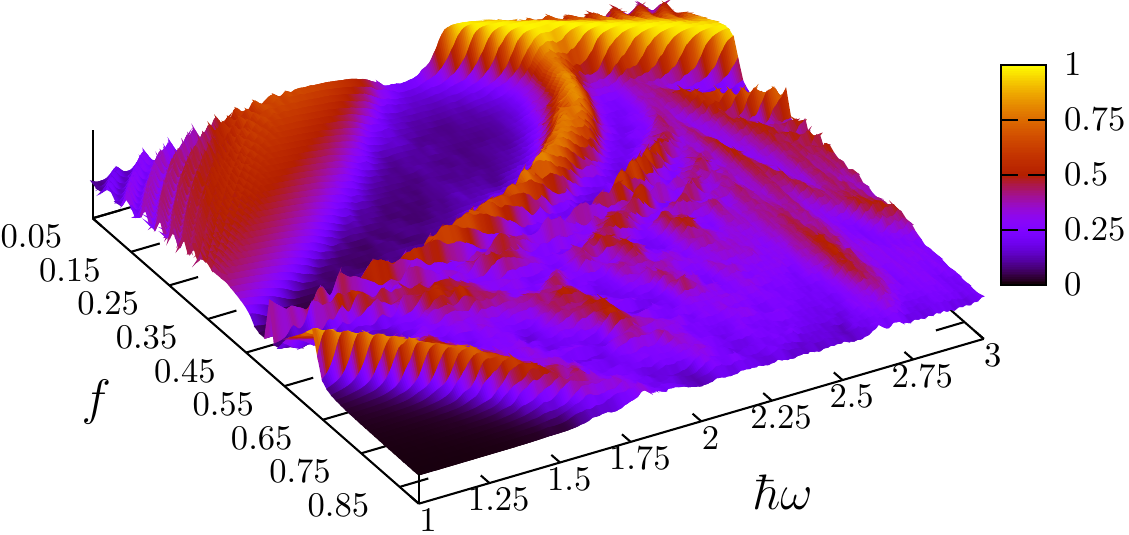}
    \caption{ \label{fig:Avsff} Absorptance of a composite slab made of
      silver spheres embedded in a dielectric matrix of $\epsilon_{\rm
        diel}=7.84$ versus $f$ and $\hbar \omega$.}
  \end{center}
\end{figure}

In this section we show that by properly combining the filling
fraction of the nanoparticles array and the dielectric constant of the
host medium, composite films can be used as tunable frequency light
absorbers.  Fig. \ref{fig:Avsff} displays the absorptance
vs. frequency of a composite slab made of silver spheres embedded in
titanium dioxide (as in the previous examples) for different values of
$f$, from $f$ = 0.05 (diluted composite of metallic spheres) up to $f$
= 0.9 (interpenetrated spheres). The slab is illuminated under normal
incidence and its thickness is 1820 nm.
 
For all the values of $f$, within the reflection band the absorptance
is less than 10\%. When $f$ increases, the spectral position of this
band shifts towards the red region of the spectrum and its bandwidth
changes.  Also, there is an absorption band at which the absorptance
is $>$ 90\%. This band is located at $\hbar \omega \approx$ 2.2 eV
(564 nm) for $f \approx$ 0.05 and its position shifts to the high
frequency region as $f$ increases, being at $\hbar \omega \approx$ 3
eV (412 nm) for $f \approx$ 0.25. From $f \approx 0.15$ up to $f$ =
0.5, there is another absorption peak at the frequency of splitting of
the reflection band that appears as a consequence of the excitation of
a multipole resonance of higher order \cite{Claro(1984)}. The spectral
position of this peak strongly depends on the value of $f$ and
redshifts as $f$ increases. For $f=\pi/6$ (close packing condition)
the system percolates and becomes conducting at low frequencies. For
larger $f$, the spheres interpenetrate and the interaction among the
dielectric voids produces an absorption band ($\approx$ 90\%) due to
the index matching between the macroscopic dielectric function and air
at NIR resonant frequencies. As $f$ increases, this band shifts to
higher frequencies, as expected.
 
\begin{figure}
  \begin{center}
    \includegraphics[width=0.9\textwidth]{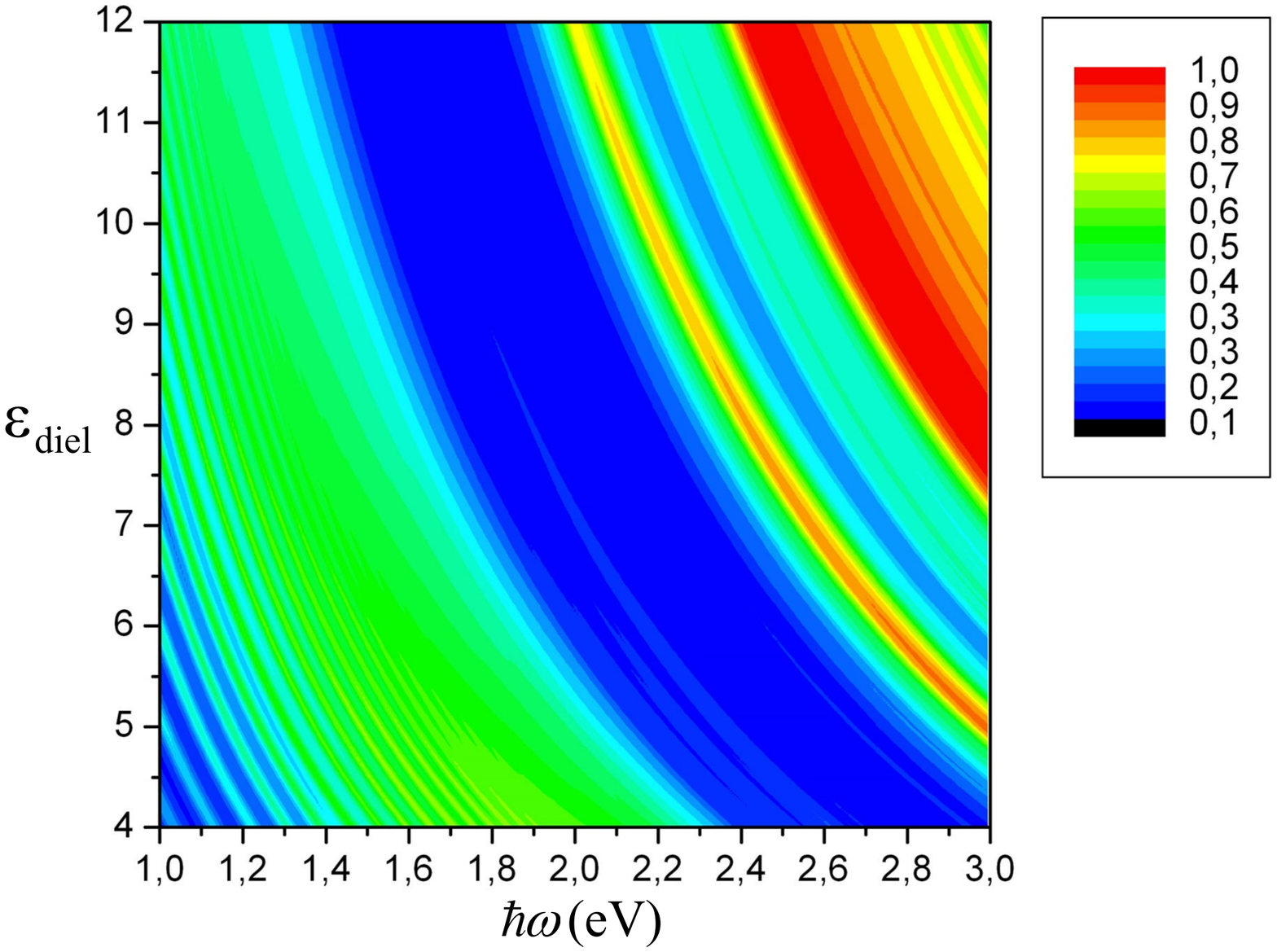}
    \caption{ \label{fig:AvsEpsh} Absorptance of a composite slab made of
      silver spheres for $f=0.23$ versus $\epsilon_{\rm diel}$ and $\hbar \omega$.}
  \end{center}
\end{figure}

So far we have investigated the absorptance of the slab by considering
different filling fractions for a fixed value of the dielectric
constant of the host medium. To investigate the influence of
$\epsilon_{\rm diel}$ in the distribution of the absorption bands, we
show in Fig. \ref{fig:AvsEpsh} the absorptance of a composite slab of
silver nanoparticles as a function of $\epsilon_{\rm diel}$ for $f$ =
0.23. As $\epsilon_{\rm diel}$ increases, the multipole resonances, and
therefore the absorption bands, shift to the red region of the
spectrum, whereas its bandwidth remains approximately constant.
 
These examples evidence that the ENR constitutes a useful and powerful
tool for the design of tunable absorbers based on metallic nanoparticle
composites: by appropriately selecting the filling fraction and the
permittivity of the host medium, the electromagnetic response can be
tuned to obtain, for example, absorption peaks and negligible
transmittance and reflectance in the visible range of the
electromagnetic spectrum.
 
\section{Conclusions}\label{conclusiones} 
 
We have considered a composite material comprising periodically
distributed metallic spheres in a dielectric host matrix. The
capabilities of the ENR approach to calculate its effective
macroscopic dielectric function have been investigated.
We have compared the optical response of a
composite slab obtained by the ENR with that of the KKR. Within the KKR framework,  
we have shown that for a fixed thickness, if the system comprises a sufficiently
large number of layers, the radius of the spheres is small enough to
guarantee the long wavelength approximation, and then a valid
comparison between both methods can be done.

Both methods adequately describe the spectral response of 
composite slabs and the qualitative agreement between the curves is
satisfactory. We analyzed in detail the behavior of the
electromagnetic response of the slab in the different frequency
regions and its relationship with the macroscopic dielectric function.
We have found characteristic features in the response such as
reflection and absorption bands whose spectral positions and widths
can be controlled by the filling fraction and the permittivities of
the particles and of the host medium. 
It is worth mentioning that although the faceted shape of the discretized particles is a
drawback for modeling spherical inclusions, it can also be viewed as an
advantage to study the optical features of synthesized metallic
nanoparticles, which exhibit facets as a consequence of the
manufacturing process~\cite{Wang(2000),Yacaman(2001),Gonzalez(2005)}. 
Then, the ENR constitutes a versatile tool to investigate the optical
properties of metallic nanoparticle composites.

\section*{Acknowledgment}

This work was supported Agencia Nacional de Promoción Científica y
Tecnológica (FONCYT-UNNE PICT-PRH-135-2008 (G.O.)), Consejo Nacional
de Investigaciones Científicas y Técnicas (PIP 112-200801-01880
(D.S. and M.I.)), Universidad de Buenos Aires (UBA-20020100100533
(D.S. and M.I.) and (UBA-20020100100327 (R.D.)), and Universidad
Nacional Autónoma de México (DGAPA-IN108413 (L.M.)).

\section*{References} 


\end{document}